# Prediction of Glass Elasticity from Free Energy Density of Topological Constraints

Collin J. Wilkinson[1], Qiuju Zheng[2,*], Liping Huang[3,*], John. C. Mauro[1,2,*]

[1]Department of Materials Science and Engineering, The Pennsylvania State University, University Park, PA 16802

[2]School of Materials Science and Engineering, Qilu University of Technology, 250353 Jinan, China

[3]Department of Materials Science and Engineering, Rensselaer Polytechnic Institute, Troy, NY 12180, USA

[*] Corresponding Authors (qlzhengqj@163.com, huangL5@rpi.edu, jcm426@psu.edu)

**Abstract**

Despite the critical importance of the elastic properties of modern materials, there is not a singular model that can predict the modulus to an accuracy needed for industrial glass design. To address this problem, we propose an approach to calculate the elastic modulus based on the free energy density of topological constraints in the glass-forming network. Our approach shows quantitatively accurate agreement with glasses across a variety of compositional families. Moreover, using temperature-dependent constraint theory, the temperature dependence of the modulus can also be predicted. Our approach is general and theoretically can be applied to any network glass.

**MAIN TEXT**

Tailoring the elastic properties of glass is extremely important in the design of new advanced compositions to control the stiffness and damage resistance of the final glass product [1–4]. Elastic moduli are defined based on the ratio of the stress exerted upon a material to the resulting strain. While different elastic moduli can be defined (e.g., Young's modulus, shear modulus, bulk modulus, and Poisson's ratio), for an isotropic material such as glass, only two of these quantities are mutually independent [1,5].

Previous attempts to model elastic moduli have focused on either computationally costly molecular dynamics simulations or empirical fitting methods. The work by Makishima and Mackenzenie [6] offered a possible solution for predicting elastic moduli based on the dissociation energy per unit volume of the glass; however, topological changes in the glass network are ignored, and the model cannot account for the temperature dependence of modulus. With molecular dynamics, elastic moduli can be obtained by applying a stress and measuring the resulting strain on the simulation cell, assuming that accurate interatomic potentials are available [3]. Machine learning has also been applied to model elastic moduli using experimentally measured composition-property databases, with a high predictive ability being achieved [3,7]. Previous analytical modeling techniques related to the topology of the glass-forming network have shown good qualitative agreement with compositional trends in modulus, but have a lack of quantitative accuracy [8–11].

Temperature-dependent constraint theory offers an alternative approach for predicting the macroscopic properties of a glass based on the underlying connectivity of its network [2,12–18]. Constraint theory considers a hierarchy of bond constraints that contribute localized rigidity to a system. In an oxide glass system, three main types of constraints are typically considered [13]:

- $\alpha$ constraints: Linear constraints between the network-forming cation (M) and the bridging oxygen (O) species.
- $\beta$ constraints: Angular constraints around the network-forming cation (O-M-O).
- $\gamma$ constraints: Angular constraints around the oxygen anion (M-O-M).

Temperature-dependent constraint theory has enabled quantitatively accurate predictions of several key properties of glass-forming systems, including fragility [13,16], glass transition temperature [13,16], hardness [2], and configurational heat capacity [15]. Constraint theory has also been used in an inverted approach to predict glass structure from the scaling of its macroscopic properties [19]. The temperature dependence of the constraints, as defined by Mauro and Gupta [13], expresses the fraction of a given type of constraint that is locally rigid at a temperature, $T$. This fraction of rigid constraints corresponds to the probability of failing to break the constraint,

$$q_n(T) = \left[1 - \exp\left(-\frac{\Delta F_n}{k_b T}\right)\right]^{vt}, \tag{1}$$

where $\Delta F_n$ is the activation free energy associated with breaking a given constraint and $vt$ is the number of escape attempts ($v$ = the vibrational attempt frequency and $t$ = the observation time). Expanding on the work of Mauro and Gupta, the energy dependence of these constraints is also a factor in governing the properties of a glass. The Wilkinson-Potter-Mauro (WPM) model [20] shows that the role of constraints is actually dynamic and is related to the potential energy of each constraint.

Glass hardness has received considerable attention in the context of topological constraint theory [2,21,22]. Several models have been proposed to explain the origin of glass hardness, all of which have a linear form

$$H_v = \frac{dH_v}{dn}[n(x) - n'], \tag{2}$$

where $H_v$ is the Vickers hardness, $x$ is chemical composition, $n'$ is the critical level of constraints needed for the substance to resist the indenter in three dimensions, and $n$ is a measure of the constraint rigidity. Recent work [2] has shown that $n$ can most accurately be defined in terms of either the density of rigid constraints (or, alternatively, the density of rigid angular constraints), given by

$$n = \frac{n_c(x)\rho(x)N_A}{M(x)}, \tag{3}$$

where $n_c$ is either the number of constraints per atom or number of angular constraints per atom, $\rho$ is the density of the composition, $M$ is the molar mass, and $N_A$ is Avogadro's number. Since hardness and elastic modulus are often correlated (hardness is the resistance to plastic deformation under indentation), one might surmise that the elastic modulus should be also some function of angular constraint density or total constraint density. However, neither of these methods can give accurate predictions of modulus.

Here we propose an improved model of elastic modulus based on the free energy density of the topological constraints,

$$\Delta F_c = \frac{\left(\Delta F_\alpha n_\alpha q_\alpha(T) + \Delta F_\beta n_\beta q_\beta(T) + \Delta F_\gamma n_\gamma q_\gamma(T)\right)\rho(x)N_A}{M(x)}, \quad (4)$$

where $q_n(T)$ is the onset function for each associated constraint to give temperature dependence to the predictions and is determined by the free energy $\Delta F_n$. The calculated values of $\Delta F_c$ is then inserted into the form of Eq. (3) and converted to Young's modulus ($E$) by

$$E = \frac{dE}{d\Delta F_c}[\Delta F_c(x) - \Delta F_c']. \quad (5)$$

Thus, the Young's modulus is controlled by the free energy parameters related to the depth of each well.

In order to determine the validity of this constraint model, Young's modulus data are collected from literature for lithium borate [6,23–34], sodium borate [6,23,32–34,24–31], and germanium selenide [35–41] glass systems. The sodium phosphosilicate [18] system was experimentally determined following the procedure described by Zheng *et al.* [42] using room temperature resonant ultrasound spectroscopy on prisms of dimensions 10 mm × 8 mm × 6 mm. Analytical topological constraint models have already been published for each of these systems [16,18,43].

For each proposed method (constraint density, angular constraint density, and free energy density of constraints), the Young's modulus was fit with the intercept and slope left as adjustable parameters ($\frac{dE}{dn}, \frac{dE}{dU_c}$). The onset temperatures for each constraint were also used as parameters in the fit of the free energy density method. Figures 1-2 show the fit for each method on phosphosilicates and sodium borates, respectively.

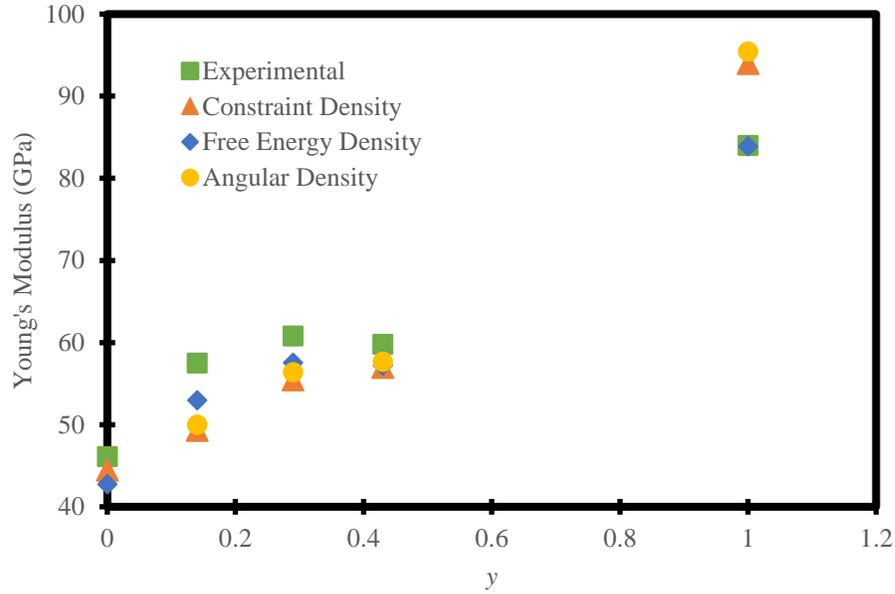

*Figure 1. The Young's modulus prediction and experimentally determined values for $0.3Na_2O \cdot 0.7(ySiO_2 \cdot (1-y)P_2O_5)$ glasses. The root mean square error of the (RMSE) values of the model predictions are: 6.41 GPa for constraint density, 3.13 GPa for free energy density, and 7.74 GPa for angular constraint density.*

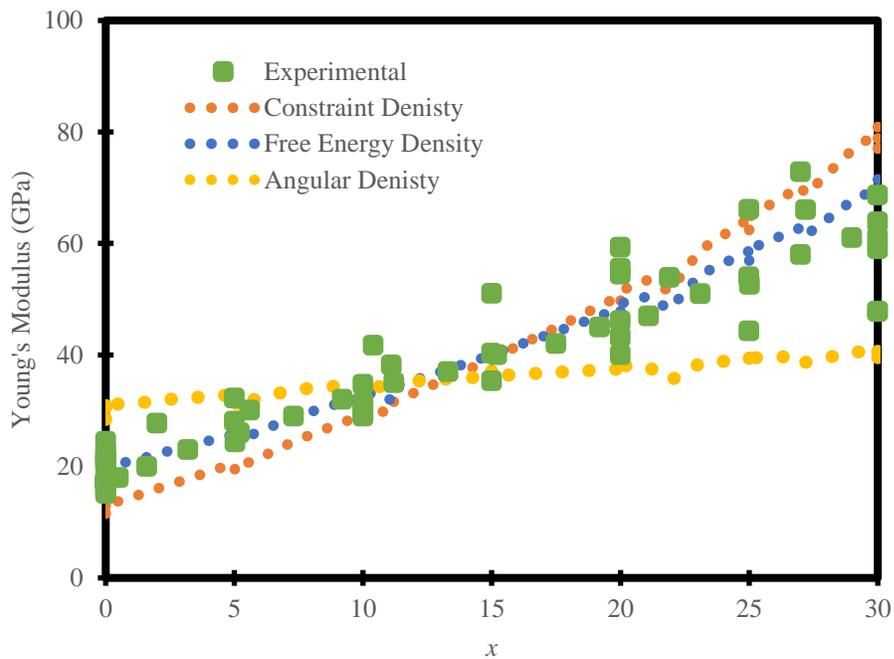

*Figure 2 The Young's modulus prediction and experimentally determined values for $xNa_2O \cdot (100-x)B_2O_3$ glasses. The RMSE values of the model predictions are: 6.35 GPa for constraint density, 4.17 GPa for free energy density, and 10.80 GPa for angular constraint density.*

We have also evaluated the model's predictive ability in terms of the temperature dependence of the Young's modulus. Using the 10% $Na_2O \cdot 90\%$ $B_2O_3$ data reported by Jaccani and Huang [1], the model and experimental predictions of the temperature dependence of the Young's modulus are plotted in Fig. 3. Here, the constraint onset temperatures are determined from fitting the room temperature data with Eqs. (5) and (6). The number of escape attempts was left as a fitting parameter for the temperature dependent predictions (the value of the escape attempts controls the width of each transition and is thermal history dependent).

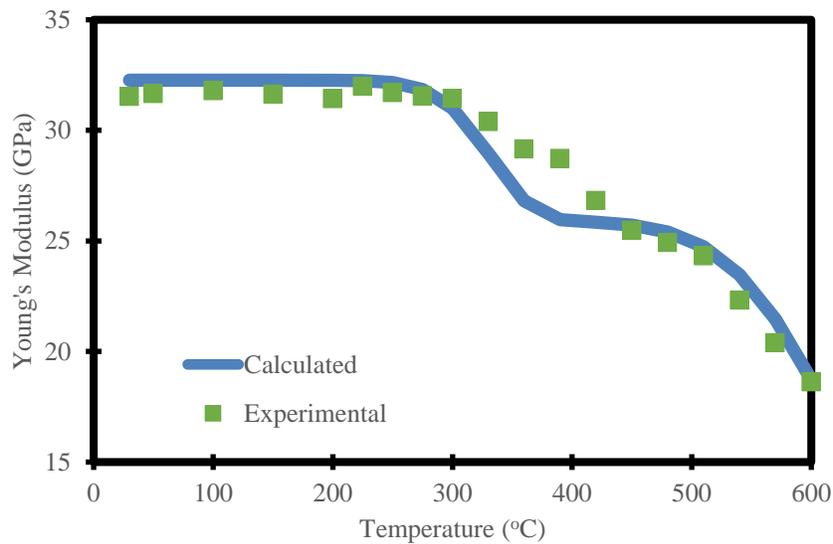

*Figure 3 Temperature dependence of the Young's modulus from theory and experiment for 10% $Na_2O$ 90% $B_2O_3$.*

It is also possible to validate this model by comparing the free parameters fitted to the Young's modulus data vs. that of previously reported data, as shown in Table 1. Results for two additional systems are plotted in Fig. 4 to show the validation of the energy density model.

Table 1 Fitted values from this analysis compared to those reported in the literature.

| Value | Fitted Value (K) | Literature (K) | Method, Citation |
|---|---|---|---|
| Silicate α Onset | 2212 | 1986 | Molecular Dynamics (Potter *et al.* [20]) |
| Silicate β Onset | 818 | 1600 | Molecular Dynamics (Potter *et al.* [20]) |
| Silicate γ Onset | 450-500 | 810 | Molecular Dynamics, (Potter *et al.* [20]) |
| Borate α Onset | 921 | Not Reported | Fitting Parameter (Mauro, Gupta, and Loucks [13]) |
| Borate β Onset | 715 | 740-760 | Fitting Parameter (Mauro, Gupta, and Loucks [13]) |
| Borate γ Onset | 393 | 328 | Fitting Parameter (Mauro, Gupta, and Loucks [13]) |

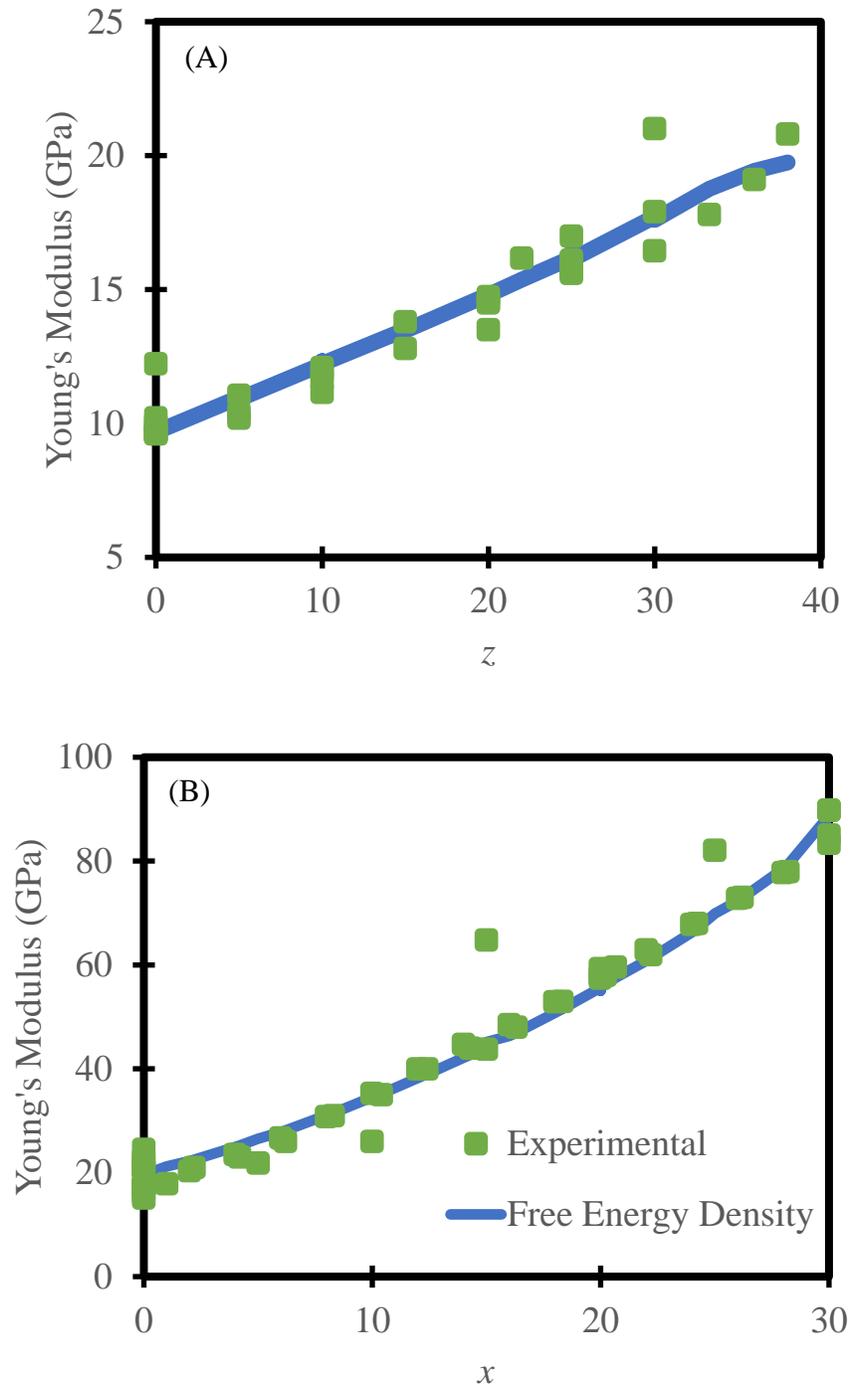

Figure 4 *The Young's modulus prediction and experimentally determined values for (A) zGe·(100-z)Se and (B) xLi$_2$O·(100-x)B$_2$O$_3$ glasses.*

In the previous work of Zheng *et al.* [2] concerning hardness of glass, the authors write,

> "It should be noted that each bond constraint corresponds to a certain energy since different kind of bond has different bonding energy… and thus the constraint density also represents an energy per unit volume. In other words, hardness is correlated to the energy per unit volume. Our findings for the borosilicate and phosphosilicate systems are further evidence in support of this argument, since both the total constraint density and angular constraint density approaches give better prediction of glass hardness compared to models based on number of atomic constraints" which makes it clear that the prediction of the hardness/modulus should be closely related to the onset temperatures (i.e., free energies) of the associated constraints. The model that was previously proposed would then perform well if all constraints had equal amounts of potential energy, but due to the drastic change in strengths between the constraints, the prediction fails.

Bauchy *et al.* [22] showed also that hardness in the calcium-silicate-hydrate system is controlled by the angular constraints, which led to the development of angular constraint density (also a metric of energy per volume) as the governing control for hardness. When the analysis of the density of angular constraints is considered through the free energy view, it becomes apparent why this method works effectively for some systems. As seen in Fig. 5, over an average coordination change of 2.5 to 4.0 the $\beta$ constraint increases nearly 50% while the other two constraints only increase 15% from their original values, meaning that most of the change comes with the variations in the $\beta$ constraint. The angular constraint about the anion, $\gamma$, is also important to consider because this constraint is more easily strained; the deviation of this angle is known to cause dramatic changes in the properties of the glass.

Models for predicting the hardness and elastic modulus of glass attempt to explicitly connect the rigid bond energy to the macroscopic properties of the system. The assumption for the hardness models is that the energies for breaking each type of constraint are all approximately equal, and hence only the number of the constraints matters. The argument can then be extended for elasticity: since elastic modulus is a bulk material property, with the approximation of all constraints being equal strength, the number of constraints per volume should be related. To correct for this approximation, when a weighted sum of free energies is used, the estimation becomes significantly more accurate. Using the density of the glass, a precise free energy per volume of the rigid constraints can be calculated. The knowledge that the energy of the bonds is tied to the elastic modulus has been widely known but had not been quantified nor placed within the context of topological constraint theory [44].

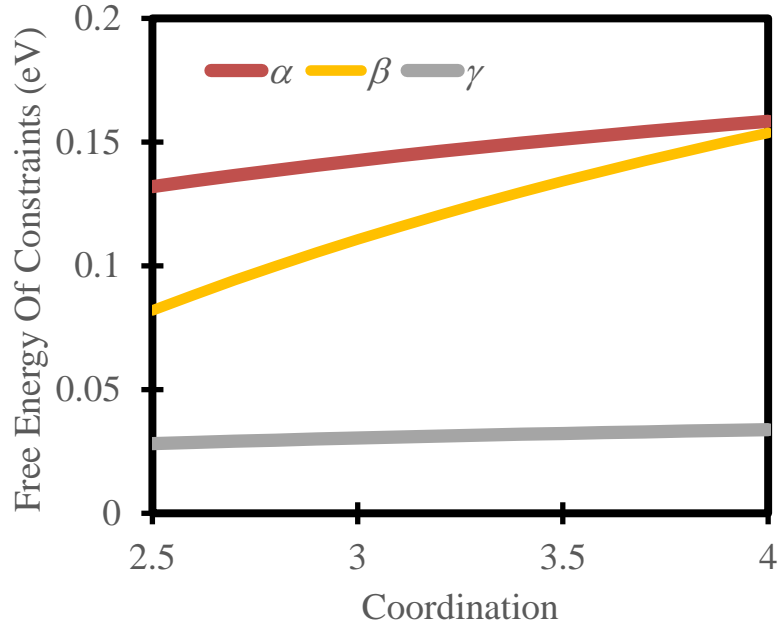

*Figure 5 The coordination of the network forming cation (assumed to be boron) and the energy changes for each individual constraint, showing that α and γ change the same percentage from their original value, but β changes dramatically more, validating the angular density approach.*

Using the temperature-dependent constraint model of Mauro and Gupta and the energy dependence of the constraints in the method laid out by Wilkinson, Potter, and Mauro [20], it is apparent that the properties are related to the free energy of the constraints in the glass network. With this approach, the elasticity of glasses can be accurately predicted across multiple compositions families. Even more remarkably, the temperature dependence of the constraints can be used to predict the modulus of a glass as a function of temperature, with the lowering modulus at high temperatures a result of the softening of the underlying topological constraints. The crux of this approach is the free energy density of the constraints, which has been shown to offer a valuable insight into the nature of constraints and properties. The validity of this approach has been demonstrated using both the compositional and temperature dependence in systems with multiple glass formers, including simple binary oxide and chalcogenide glasses, with excellent agreement in all cases. Moreover, the constraint onset temperatures are found to be in close agreement with values found in the literature.

**Acknowledgments:**

Special thanks to Arron R. Potter, Rebecca S. Welch, and Karan Doss for many insightful conversations. C.J. Wilkinson and J.C. Mauro are grateful for financial support from Corning Incorporated. Q. Zheng would like to acknowledge support by National Natural Science Foundation of China (51802165) and Shandong Provincial Natural Science Foundation, China (ZR2017LEM007). L. Huang acknowledges the financial support from the US National Science Foundation under grant No. DMR-1255378 and DMR-1508410.